\documentclass[prd,aps,reprint,noeprint,superscriptaddress,amsmath, amssymb]{revtex4-1}
\usepackage{graphicx}
\usepackage{subfigure}
\usepackage{lineno}
\usepackage{epsfig}
\usepackage{multirow}
\usepackage{overpic}
\usepackage{color}
\usepackage{bm} 
\usepackage{xspace} 
\usepackage{dcolumn}
\usepackage{booktabs,tabularx}
\usepackage{array}
\usepackage{textcomp}
\usepackage[colorlinks,linkcolor=blue,urlcolor=blue,citecolor=blue]{hyperref}
\usepackage{tikz}

\usepackage{cleveref}

\newcommand{\BESIIIorcid}[1]{\href{https://orcid.org/#1}{\hspace*{0.1em}\raisebox{-0.45ex}{\includegraphics[width=1em]{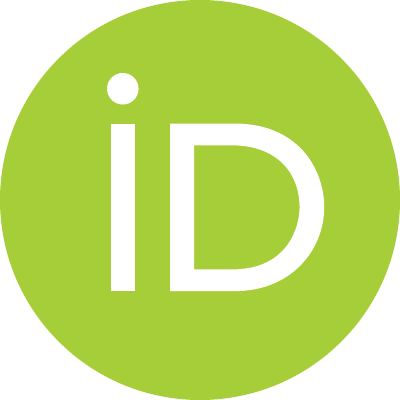}}}}
\begin{document}
\normalsize
\parskip=5pt plus 1pt minus 1pt


\title{
\boldmath First measurement of the absolute branching fractions of $\Sigma^+$ nonleptonic decays and test of the $\Delta I = 1/2$ rule
}

\author{
M.~Ablikim$^{1}$\BESIIIorcid{0000-0002-3935-619X},
M.~N.~Achasov$^{4,c}$\BESIIIorcid{0000-0002-9400-8622},
P.~Adlarson$^{81}$\BESIIIorcid{0000-0001-6280-3851},
X.~C.~Ai$^{86}$\BESIIIorcid{0000-0003-3856-2415},
R.~Aliberti$^{39}$\BESIIIorcid{0000-0003-3500-4012},
A.~Amoroso$^{80A,80C}$\BESIIIorcid{0000-0002-3095-8610},
Q.~An$^{77,64,\dagger}$,
Y.~Bai$^{62}$\BESIIIorcid{0000-0001-6593-5665},
O.~Bakina$^{40}$\BESIIIorcid{0009-0005-0719-7461},
Y.~Ban$^{50,h}$\BESIIIorcid{0000-0002-1912-0374}
% ...
\\
\vspace{0.2cm}
(BESIII Collaboration)\\
\vspace{0.2cm} {\it
$^{1}$ Institute of High Energy Physics, Beijing 100049, People's Republic of China\\
% ...
$^{4}$ Budker Institute of Nuclear Physics SB RAS (BINP), Novosibirsk 630090, Russia\\
% ...
$^{39}$ Johannes Gutenberg University of Mainz, Johann-Joachim-Becher-Weg 45, D-55099 Mainz, Germany\\
$^{40}$ Joint Institute for Nuclear Research, 141980 Dubna, Moscow region, Russia\\
% ...
$^{50}$ Peking University, Beijing 100871, People's Republic of China\\
% ...
$^{62}$ Southeast University, Nanjing 211100, People's Republic of China\\
% ...
$^{64}$ State Key Laboratory of Particle Detection and Electronics, Beijing 100049, Hefei 230026, People's Republic of China\\
% ...
$^{77}$ University of Science and Technology of China, Hefei 230026, People's Republic of China\\
% ...
$^{80}$ University of Turin and INFN, (A)University of Turin, I-10125, Turin, Italy; (B)University of Eastern Piedmont, I-15121, Alessandria, Italy; (C)INFN, I-10125, Turin, Italy\\
$^{81}$ Uppsala University, Box 516, SE-75120 Uppsala, Sweden\\
% ...
$^{86}$ Zhengzhou University, Zhengzhou 450001, People's Republic of China\\
\vspace{0.2cm}
$^{\dagger}$ Deceased\\
% ...
$^{c}$ Also at the Novosibirsk State University, Novosibirsk, 630090, Russia\\
% ...
$^{h}$ Also at State Key Laboratory of Nuclear Physics and Technology, Peking University, Beijing 100871, People's Republic of China\\
% ...
}
}
\begin{abstract}
Based on $(10087 \pm 44) \times 10^6$ $J/\psi$ events collected by the BESIII detector at the center-of-mass energy $\sqrt{s} = 3.097$ GeV,
the first absolute measurement of the branching fractions for the decays $\Sigma^+ \to p \pi^0$ and $\Sigma^+ \to n \pi^+$ is performed.
The branching fractions are determined to be $B_{\Sigma^+ \to p \pi^0} = (49.79 \pm 0.06 \pm 0.22)\%$ and $B_{\Sigma^+ \to n \pi^+} = (49.87 \pm 0.05 \pm 0.29)\%$, where the first uncertainties are statistical and the second systematic.
Our results show significant deviations from the PDG values for both decays,
with differences of 4.4$\sigma$ for $\Sigma^+ \to p \pi^0$ and 3.4$\sigma$ for $\Sigma^+ \to n \pi^+$. 
These measurements will play important roles in precise studies of the decay properties of strange, charmed, and beauty baryons 
with $\Sigma^+$ in the final states.
Furthermore, the $\Delta I = 1/2$ rule is tested in nonleptonic $\Sigma^\pm$ decays.
The observed results deviate from zero by more than $5\sigma$, indicating the presence of the $\Delta I = 3/2$ transition amplitude in the $\Sigma$ hyperon decays.
\end{abstract}

\maketitle

The study of hyperons has long provided a unique window into the non-perturbative regime of the Standard Model (SM),
where the weak decays of strange baryons serve as sensitive probes of hadronic structure and fundamental
symmetries~\cite{Lee:1957qs,Donoghue:1985ww,Jenkins:1991bt,Cabibbo:2003ea,Li:2019knv,Goudzovski:2022vbt,Shi:2025xkp, Perotti:2018wxm, He:2025ibc}.
In recent years, both collider and fixed-target experiments have driven a renewed surge of interest in hyperon physics.
In particular, the BESIII experiment has performed numerous studies of hyperon production and decay,
exploiting quantum-entangled hyperon–antihyperon pairs produced in $J/\psi$ and $\psi(3686)$ decays~\cite{BESIII:2018cnd,BESIII:2020fqg,BESIII:2021ypr,BESIII:2022qax,BESIII:2022rgl,BESIII:2023sgt,BESIII:2023drj,BESIII:2023ldd,BESIII:2023fhs,BESIII:2023clq,BESIII:2024nif,BESIII:2024geh,BESIII:2024wvw,BESIII:2024lio,BESIII:2025jxt}.
These developments demonstrate that precise experimental measurements of hyperon decays provide powerful probes of the interplay between weak and strong interactions in the nonperturbative regime of the SM.

Recently, the absolute branching fraction (BF) of $\Sigma^+ \to p\gamma$
($\mathcal{B}_{\Sigma^+ \to p\gamma}$) was measured for the first
time using the double-tag method by BESIII~\cite{BESIII:2023fhs},
revealing a discrepancy of 4.2 standard deviations from the
Particle Data Group (PDG) value~\cite{PDG}. Notably, this PDG value
was not derived from a direct measurement of the absolute BF, but
inferred by multiplying the BF of \mbox{$\Sigma^+ \to p\pi^0$}
($\mathcal{B}_{\Sigma^+ \to p\pi^0}$) with the ratio
$\mathcal{B}_{\Sigma^+ \to p\gamma}/\mathcal{B}_{\Sigma^+ \to
p\pi^0}$, which was averaged from several measurements conducted before 1995~\cite{E761:1994yxs, Hessey:1989ep, Kobayashi:1987yv}. Moreover, the PDG value of $\mathcal{B}_{\Sigma^+ \to
p\pi^0}$ itself was determined from various relative measurements
with missing systematic uncertainties.
The significant discrepancy observed by BESIII then raises questions
about the reliability of the current PDG value of
$\mathcal{B}_{\Sigma^+ \to p\pi^0}$, and a recent theoretical paper predicts a value for $\mathcal{B}_{\Sigma^+ \to p\pi^0}$ that is $1.5\sigma$ below 
the PDG average~\cite{Wu:2025hnh}, 
thereby highlighting the need for
an independent and direct absolute measurement. 
Besides, the PDG value for the BF of $\Sigma^+ \to n \pi^+$
($\mathcal{B}_{\Sigma^+ \to n\pi^+}$) also relies solely on relative
measurements, further calling for the absolute BF measurement of
these decays.
Precise measurements of these decay modes are essential for detailed studies of the decay properties of many other baryons--such as the 
$\Lambda(1405)$, $\Sigma(1385)^+$, $\Lambda_c$, $\Xi_c$-- that decay into final states involving $\Sigma^+$.

Although isospin symmetry is explicitly broken in weak interactions, experiments have established a striking empirical pattern known as the $\Delta I = 1/2$ rule, which describes the predominance of transition amplitudes with a total isospin change of $1/2$ in nonleptonic strange-hadron decays~\cite{Cirigliano:2011ny,HFLAV:2022esi}. This rule has been well verified in both kaon and hyperon systems. 
Lattice QCD calculations have provided a first-principles derivation of this enhancement in $K\to\pi\pi$ decays, 
demonstrating that the observed hierarchy between the $I=1/2$ and $I=3/2$ amplitudes originates from nonperturbative QCD dynamics~\cite{RBC:2015gro}. However, recent observations of deviations from this rule in $\Omega^-$ and $\Lambda$ hyperon decays~\cite{BESIII:2023ldd,BESIII:2023jhj} raise important questions about its universality. These developments motivate further experimental tests of
the $\Delta I = 1/2$ rule in hyperon decays.
In $\Sigma$ hyperon decays, the $\Delta I =1/2$ rule imposes specific constraints on the decay amplitudes~\cite{Donoghue:2022wrw} 
\begin{equation}
    \label{eq:testDeltaI}
    {\cal M}_{\Sigma^-\rightarrow n\pi^-} - {\cal M}_{\Sigma^+ \rightarrow n\pi^+} + \sqrt{2}{\cal M}_{\Sigma^+\rightarrow p\pi^0} = 0,
\end{equation}
where $\cal M$ represents the $A$ or $B$ amplitudes~\cite{Donoghue:2022wrw}, 
corresponding to the parity-violating $S$-wave and parity-conserving $P$-wave components of these decays~\cite{PDG}, respectively.
The $A$, $B$ (also the $S$-wave, and $P$-wave) amplitudes can be extracted from the precisely measured BFs 
and the corresponding decay parameters (see appendix~\ref{sec:A1} and appendix~\ref{sec:A2} for details). 
With the updated $A$ and $B$ amplitudes of
the $\Sigma^+$ nonleptonic decays,
a new and stringent test of the $\Delta I= 1/2$ rule can be performed. 
Moreover, the extracted $S$-wave and $P$-wave amplitudes 
provide critical insights into the longstanding $S$-wave/$P$-wave puzzle~\cite{Donoghue:1985dya},
and are indispensable for theoretical predictions of both $CP$ violation in these decay modes~\cite{He:2025ibc} 
and the rare decays $\Sigma^+ \to p l^+ l^-$ ($l=e$ or $\mu$)~\cite{Roy:2024hqg}.

The large yield of quantum-entangled $\Sigma^+ \bar{\Sigma}^-$ pairs
at BESIII~\cite{Li:2016tlt} enables the first precise measurement of the
absolute BFs of the dominant $\Sigma^+$ non-leptonic decays, $\Sigma^+ \to p \pi^0$ and $\Sigma^+ \to n \pi^+$, by employing
a double-tag (DT) method~\cite{MARK-III:1985hbd}.
The single-tag (ST) events of $\bar{\Sigma}^-$ hyperons are reconstructed via the decay $\bar{\Sigma}^- \to \bar{p} \pi^0$.
The events where a signal candidate can be reconstructed from the particles recoiling against the ST $\bar{\Sigma}^-$ hyperons are called DT events.
To improve the detection efficiencies, only the $p$ and $\pi^+$ from $\Sigma^+$ decays are reconstructed on the signal side for the decay modes $\Sigma^+ \to p \pi^0$ and $\Sigma^+ \to n \pi^+$, respectively,
whereas the $\pi^0$ and the neutron are not reconstructed (unless otherwise noted, the charge-conjugated decays are always implied).
The BF of a signal decay is determined by
\begin{equation}
    \label{eq:BFcalcu}
    \mathcal{B}_{\rm sig} = \frac{{N}_{\rm DT}~\epsilon_{\rm ST}}{{N}_{\rm ST}~\epsilon_{\rm DT}} = \frac{{N}_{\rm DT}}{{N}_{\rm ST}~\epsilon_{\rm sig}},
\end{equation}
where ${N}_{\rm ST}$ and ${N}_{\rm DT}$ are the ST and DT yields, respectively.
The $\epsilon_{\rm sig} = \epsilon_{\rm DT}/\epsilon_{\rm ST}$ is the signal efficiency in the presence of
an ST $\bar{\Sigma}^-$ hyperon, with $\epsilon_{\rm ST}$ and $\epsilon_{\rm DT}$ denoting the ST and DT efficiencies.

The analysis presented in this Letter is based on a sample of
$(10087 \pm 44) \times 10^6$ $J/\psi$ events~\cite{BESIII:2021cxx} collected
with the BESIII detector at the BEPCII collider.  Details about BEPCII
and BESIII can be found in Refs.~\cite{Ablikim2010, BESIII:2020nme,
  Yu:2016cof, Li:2017eToF, Guo:2017eToF, Cao:2020ibk}.
Simulated data samples produced with  {\sc geant4}-based~\cite{G42002iii} Monte Carlo (MC) software,
which includes the geometric description of the BESIII detector and the
detector response, are used to determine detection efficiencies and
estimate background contribution. The simulations model the beam energy spread
and initial state radiation in the $e^+e^-$ annihilations with
the generator {\sc kkmc}~\cite{Jadach2001}.  Inclusive MC sample
of $J/\psi$ resonance is produced, in which the
known decay modes are modeled with {\sc evtgen}~\cite{Lange2001,*Ping2008},
and the remaining unknown decays are modeled with {\sc lundcharm}~\cite{Chen2000,*Yang2014a}.
To determine the detection efficiencies, the signal MC samples,
\mbox{$ J/\psi \to \Sigma^{+} (\to p \pi^0) \bar{\Sigma}^- (\to \bar{p} \pi^0)$},
\mbox{$J/\psi \to \Sigma^{+} (\to p \pi^0) \bar{\Sigma}^- (\to \bar{n} \pi^-) +c.c.$},
are generated according to the angular distributions measured in Refs.~\cite{BESIII:2025jxt, BESIII:2023sgt}.
In addition, a phase space MC sample of $J/\psi \to \Delta^{+} \bar{\Delta}^-, \Delta^{+} \to$ $anything$
is generated to study potential backgrounds, where the $\Delta^+$ mass and width are fixed at the PDG value.

Charged tracks detected in the multilayer drift chamber (MDC) are required to be within a polar angle
($\theta$) range of $|\rm{cos\theta}|<0.93$, where $\theta$ is defined with respect to the $z$-axis,
which is the symmetry axis of the MDC. For each track,
the point of closest approach to the interaction point must be within $2$~cm in the plane perpendicular to the beam direction
and within $10$~cm along the beam direction.
Particle identification (PID) for charged tracks combines measurements of the specific ionization energy loss in the MDC ($\mathrm{d}E/\mathrm{d}x$)
and the flight time in the time-of-flight system to form likelihoods $\mathcal{L}(h)(h = p, K, \pi)$ for various hadron $h$ hypotheses.
Charged tracks with $\mathcal{L}(p)>\mathcal{L}(K)$, $\mathcal{L}(p)>\mathcal{L}(\pi)$ and $\mathcal{L}(p)>0.001$ are
identified as protons, and those with
$\mathcal{L}(\pi)>\mathcal{L}(K)$ are identified as pions.

The photons used to reconstruct the $\pi^0$ candidates are identified using showers in the electromagnetic calorimeter~(EMC).
The deposited energy of each shower must be more than 25~MeV in the barrel region~(\mbox{$|\cos\theta| < 0.80$})
and more than 50~MeV in the end-cap region~(\mbox{$0.86<|\cos\theta|<0.92$}).
To exclude showers that originate from charged tracks, the angle subtended by the EMC shower
and the position of the closest charged track at the EMC must be greater than 10 degrees (20 degrees if the track is anti-proton)
as measured from the interaction point.
To suppress electronic noise and showers unrelated to the event,
the difference between the EMC time of the photon candidate and the event start time is required to be within [0, 700]~ns.
The $\pi^0$ candidates are formed by the $\gamma \gamma$ pairs,
with the requirements that at least one photon comes from the barrel region and the invariant mass is within the range
$[0.115, 0.150]$~GeV/$c^2$.
A kinematic fit constraining the $\gamma \gamma$ invariant mass to the known $\pi^0$ mass~\cite{PDG} is performed,
and the updated four-momentum is used in further analysis.

The ST $\bar{\Sigma}^-$ candidates are required to have at least one $\bar{p}$ and one $\pi^0$.
If there is more than one $\bar{p} \pi^0$ candidate,
the one with the minimum $|M_{\bar{p} \pi^0} - m_{\bar{\Sigma}^-}|$ is retained for further analysis,
where $m_{\bar{\Sigma}^-}$ is the known $\bar{\Sigma}^-$ mass~\cite{PDG}.
The invariant mass of the retained $\bar{p} \pi^0$ pair is required to be within $[1.169, 1.205]$~GeV/$c^2$.

\begin{figure}[htbp]
    \begin{center}
        \begin{overpic}[width = 0.8\linewidth]{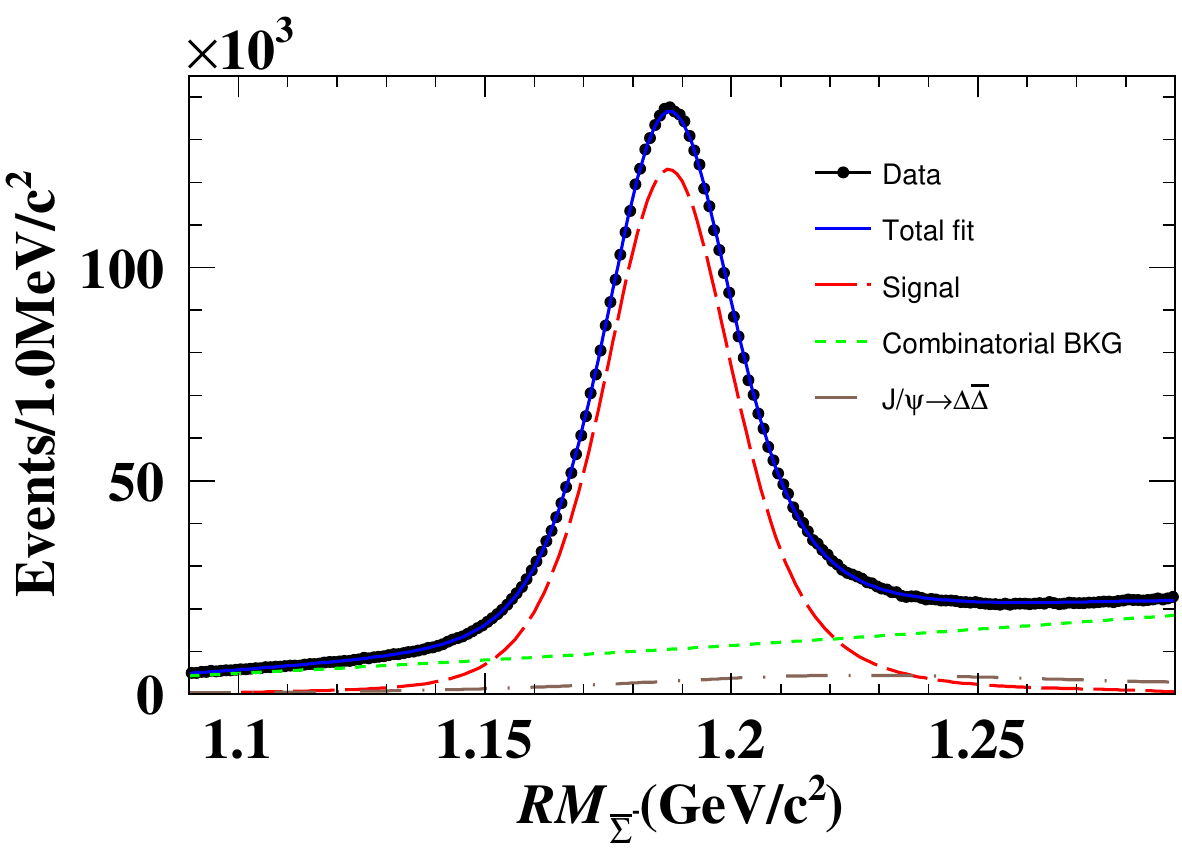}
        \end{overpic}
    \end{center}
    \caption{Fit to the $RM_{\bar{\Sigma}^-}$ distribution for the ST sample.
    The dots with error bars are data, the blue solid line is the total fit,
    the red long-dashed line denotes the fitted signal shape,
    and the green short-dashed and brown dot-dashed lines represent the combinatorial background and
    $J/\psi \to \Delta^{+} \bar{\Delta}^-$ shapes, respectively.
    }
    \label{fig:SingleTag}
\end{figure}

After applying the above selection,
the ST yield is determined by performing an unbinned maximum likelihood fit
to the recoil-mass spectrum of the reconstructed $\bar{\Sigma}^-$ candidate, shown in Fig.~\ref{fig:SingleTag},  and defined as
\begin{equation}
RM_{\bar{\Sigma}^-} =
    \sqrt{\frac{(E_{cms}-E_{\bar{p} \pi^0})^2}{c^4} - \frac{\vec{P}_{\bar{p}\pi^0}^2}{c^2}} + M_{\bar{p}\pi^0}- m_{\bar{\Sigma}^-}.
    \label{Eq:mRec}
\end{equation}
Here, $E_{cms}$ is the center-of-mass energy, $E_{\bar{p} \pi^0}$ and $\vec{P}_{\bar{p} \pi^0}$
are the energy and momentum of the tagged $\bar{p} \pi^0$ in the $J/\psi$ rest frame, respectively.
The term $M_{\bar{p}\pi^0}- m_{\bar{\Sigma}^-}$ is introduced to improve the recoil-mass resolution of $\bar{\Sigma}^-$.
In the fit, the signal shape is described by the MC-simulated shape convolved with a Gaussian function,
where the Gaussian function is used to compensate for the difference
in the mass resolution between data and MC simulation.
The background shape, which is dominated by combinatorial backgrounds and a broad peak from $J/\psi \to \Delta^{+} \bar{\Delta}^-$,
is described by the MC-simulated shape from $J/\psi \to \Delta^{+} \bar{\Delta}^-$ plus a second-order Chebyshev polynomial.
The ST yield is determined to be $4504566 \pm 3320$, with a corresponding efficiency of $41.66\%$, as listed in Table~\ref{tab:BFs}.

\begin{table*}[htbp]
    \caption{The ST yields ($\rm{N}_{\rm ST}$), ST efficiency
    ($\epsilon_{\rm ST}$), DT yields ($\rm{N}_{\rm DT}$), DT
    efficiency ($\epsilon_{\rm DT}$) and the absolute BFs of the two
    dominant $\Sigma^+$ decays, $\Sigma^+ \to p \pi^0$ and $\Sigma^+ \to n \pi^+$.
    The uncertainties are statistical only.}
    \label{tab:BFs}
    \setlength{\extrarowheight}{1.0ex}
     \renewcommand{\arraystretch}{1.1}
    \begin{center}
        \scalebox{1.0}{
        \begin{tabular} {l | c | c | c | c | c}
            \hline \hline
            Decay mode                                     & $\rm{N}_{\rm ST}$ & $\epsilon_{\rm ST} (\%)$ & $\rm{N}_{\rm DT}$ & $\epsilon_{\rm DT} (\%)$ & BF (\%)    \\
            \hline
            $\Sigma^+ \to p \pi^0$ & \multirow{2}{2.5cm}{\centering $4504566 \pm 3320$} & $\multirow{2}{1.2cm}{\centering 41.66}$ & $~1437863\pm1672~$ & $~26.71~$ & $~49.79\pm 0.06~$\\
            \cline{1-1} \cline{4-6}
            $\Sigma^+ \to n \pi^+$ & &  & $1638422 \pm 1780$ & $30.38$ & $49.87\pm0.05$\\
            \hline \hline
        \end{tabular}}
    \end{center}
\end{table*}

The signal processes $\Sigma^+ \to p \pi^{0}$ and $\Sigma^+ \to n \pi^{+}$ are searched for in the remaining charged tracks
on the recoil side of the ST $\bar{\Sigma}^-$ candidates. As aforementioned, only the charged tracks $p$ and $\pi^+$ are
reconstructed for the decays $\Sigma^+ \to p \pi^{0}$ and $\Sigma^+ \to n \pi^{+}$, respectively.
To suppress background contributions, the number of additional charged tracks in the event is required to be zero.
In the BF measurement of $\Sigma^+ \to p \pi^{0}$, if both $\Sigma^+$ and $\bar{\Sigma}^-$ can be reconstructed in one event,
this event would be counted twice in the DT sample ($\Sigma^+ \bar{p}$ and $\bar{\Sigma}^- p$).
To avoid double counting, only the combination with the tagged $\Sigma$ that has the closest invariant mass to the $\Sigma^+$ known mass~\cite{PDG}
is kept as a DT event for further study.

The recoil-masses against the reconstructed $\bar{\Sigma}^- p$ ($\bar{\Sigma}^- \pi^+$), defined as
\begin{align}
    \label{eq:RMSigmaX}
    RM_{\bar{\Sigma}^-p(\pi)} &= \sqrt{\frac{(E_{cms}-E_{\bar{p} \pi^0} - E_{p(\pi)})^2}{c^4} - \frac{(\vec{P}_{\bar{p} \pi^0} + \vec{P}_{p(\pi)})^2}{c^2}} \notag\\
    &+ M_{\bar{p}\pi^0}- m_{\bar{\Sigma}^-},
\end{align}
accumulate around the invariant mass of the un-reconstructed particle $\pi^0$ ($n$).
Here, $E_{p(\pi)}$ and $\vec{P}_{p(\pi)}$ are the energy and momentum of the $p(\pi^+)$ in the $J/\psi$ rest frame, respectively.
The $RM_{\bar{\Sigma}^-p(\pi)}$ is used to suppress the background events by requiring it
to lie within $[0.034, 0.231] ([0.881, 0.997])$~GeV/$c^2$, corresponding to $\pm 3\sigma$ around the known $\pi^0(n)$ mass,
where $\sigma$ denotes the resolution of the corresponding recoil-mass, the distribution of $RM_{\bar{\Sigma}^-p(\pi)}$ can be found in Ref.~\ref{sec:A3}.
Then, the DT yield is extracted by performing an unbinned maximum
likelihood fit using the same fitting strategy as described
for the ST sample. The signal and background shapes are taken
from the MC samples corresponding to the DT selection.
The number of DT events of the $\bar{\Sigma}^- p (\bar{\Sigma}^- \pi^+)$ sample is determined to be
$1437863\pm1672(1638422 \pm 1780)$,
with a corresponding DT efficiency of $26.71(30.38) \%$, as summarized in Table~\ref{tab:BFs}.


\begin{figure}[htbp]
    \begin{center}
        \begin{overpic}[width = 1.0\linewidth]{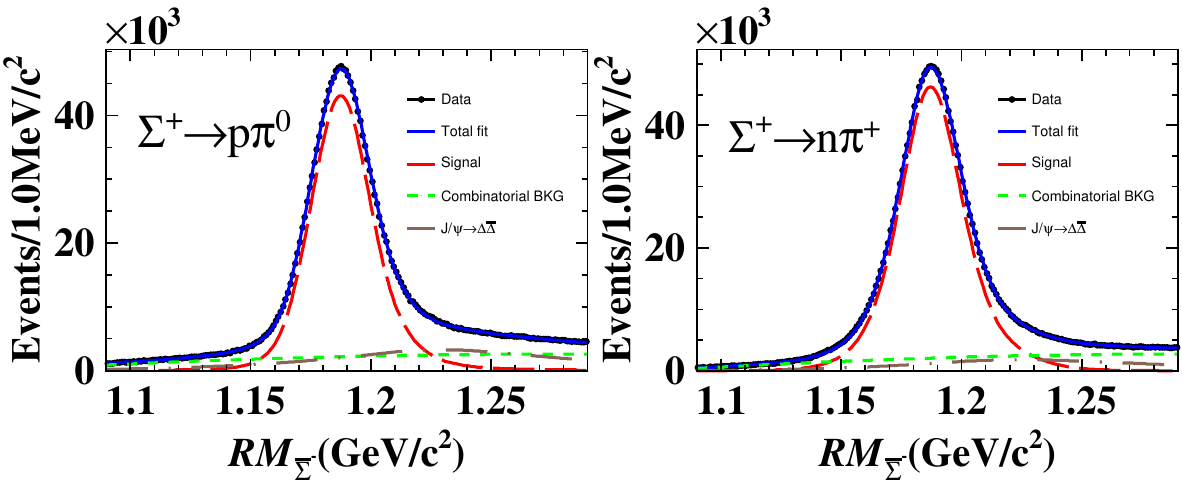}
        \end{overpic}
    \end{center}
    \caption{Fit to the $RM_{\bar{\Sigma}^-}$ distributions for the DT $\bar{\Sigma}^- p$ (left)
    and $\bar{\Sigma}^- \pi^+$ (right) samples.
    The dots with error bars are data, the blue solid line is the total fit,
    the red long-dashed line denotes the fitted signal shape,
    and the green short-dashed and brown dot-dashed lines represent the combinatorial background and
    $J/\psi \to \Delta^{+} \bar{\Delta}^-$ shapes, respectively.
    }
    \label{fig:DT}
\end{figure}

Based on Eq.~\ref{eq:BFcalcu}, the absolute BFs of $\Sigma^+ \to p \pi^0$ and $\Sigma^+ \to n \pi^+$
are determined to be $\mathcal{B}_{\Sigma^+ \to p \pi^0} = (49.79\pm0.06)\%$ and $\mathcal{B}_{\Sigma^+ \to n \pi^+} = (49.87 \pm
0.05)\%$, as listed in Table~\ref{tab:BFs}.
The uncertainties here are only statistical.

With the DT method, most systematic uncertainties associated with the ST selection cancel out.
The sources of the systematic uncertainties are summarized in Table~\ref{tab:systematic}, and each contribution is discussed below.

\begin{table}[htbp]
    \caption{Relative systematic uncertainties of BF measurement in \%.}
    \label{tab:systematic}
    \setlength{\extrarowheight}{1.0ex}
     \renewcommand{\arraystretch}{1.0}
    \begin{center}
    \scalebox{1.0}{
        \begin{tabular} {l c c}
            \hline \hline
            Source & $\mathcal{B}_{\Sigma^+ \to p \pi^0}$ & $\mathcal{B}_{\Sigma^+ \to n \pi^+}$\\
            \hline
            Proton/pion tracking                                                & $0.14$ & $0.27$      \\
            Proton/pion PID                                                     & $0.09$ & $0.33$      \\
            Signal Shape                                                        & $0.18$ & $0.26$   \\
            Background Shape                                                    & $0.30$ & $0.26$   \\
            BKG fluctuation of ST yields                                        & $0.06$ & $0.06$   \\
            DT peaking BKG                                                      & $0.12$ & $-$   \\
            Misidentification of $p/\pi$                                       & $0.01$ & $0.05$   \\
            Requirement on $RM_{\bar{\Sigma}^- p}/M_{\bar{\Sigma}^- \pi^+}$     & $0.14$ & $0.16$   \\
            MC generator                                                        & $0.05$ & $0.04$   \\
            MC statistics                                                       & $0.04$ & $0.03$   \\
            \hline \hline
            Total                                                               & $0.44$ & $0.59$     \\
            \hline \hline
        \end{tabular}
    }
    \end{center}
\end{table}

The systematic uncertainties arising from the $p,~\pi^+$ tracking and PID efficiencies are estimated by a
control sample of $J/\psi \to p \bar{p} \pi^+ \pi^-$.
The efficiency differences between data and MC simulation from the control samples are
used to reweight the signal MC samples. The differences between the detection efficiencies with and without corrections are taken as the systematic uncertainties, which are 0.14\% (0.09\%) for $\mathcal{B}_{\Sigma^+ \to p \pi^0}$ measurement
and 0.27\% (0.33\%) for $\mathcal{B}_{\Sigma^+ \to n \pi^+}$, corresponding to the tracking (PID) contributions.

The systematic uncertainties related to the ST and DT signal shapes of the $RM_{\bar{\Sigma}^-}$ distribution are estimated
by varying the signal MC shape convolved with a Gaussian function in the fit model to the signal MC shape. The resulting changes
in the BFs are taken as the systematic uncertainties, which are 0.18\% and 0.26\% for $\mathcal{B}_{\Sigma^+ \to p \pi^0}$
and $\mathcal{B}_{\Sigma^+ \to n \pi^+}$, respectively.

The uncertainties associated with the ST and DT background shapes are estimated by refitting the distributions with alternative fourth-order Chebyshev polynomials to describe all background components due to the relative broad distribution
of the $J/\psi \to \Delta^+ \bar{\Delta}^-$ process in the fit range.
The resulting uncertainties are 0.30\% and 0.26\% for
$\mathcal{B}_{\Sigma^+ \to p \pi^0}$ and $\mathcal{B}_{\Sigma^+ \to n \pi^+}$, respectively.
In addition, the uncertainty due to the background statistical fluctuation of the ST yield, 0.06\%,
is also considered as a systematic uncertainty.

In the analysis, the $RM_{\bar{\Sigma}^-}$ distributions are used to
determine the DT yields.  All the $\Sigma^+$ decay channels with a
proton in the final state that can pass the event selection are
referred to as the signal in the fit for DT sample of
$\bar{\Sigma}^- p$, and the channels with a $\pi^+$ are regarded as
the signal for $\bar{\Sigma}^- \pi^+$.  Therefore, MC simulations of
all the known $\Sigma^+$ decay modes other than the signal processes
with a proton or $\pi^+$, namely $\Sigma^+\to p \gamma$,
$\Sigma^+ \to \Lambda e^+ \nu_e$, and $\Sigma^+ \to n\pi^+\gamma$, are
generated to estimate the potential impact on the BFs from these
channels.  These uncertainties (denoted as DT peaking BKG in
Table~\ref{tab:systematic}) are evaluated to be $0.12\%$ for
$\mathcal{B}_{\Sigma^+ \to p \pi^0}$ and negligible for
$\mathcal{B}_{\Sigma^+ \to n \pi^+}$.  The misidentifications of
proton and pion could also contribute to the systematic uncertainties,
and they are determined to be 0.01\% and 0.05\% for
$\mathcal{B}_{\Sigma^+ \to p \pi^0}$ and
$\mathcal{B}_{\Sigma^+ \to n \pi^+}$, respectively, from the MC simulation.

The mass windows of $RM_{\bar{\Sigma}^- p}$ and $RM_{\bar{\Sigma}^- \pi^+}$ distributions
are about $\pm 3\sigma$ around the known $\pi^0$ and neutron masses~\cite{PDG}, respectively.
We change the mass window to
$\pm 2.5\sigma$, $\pm 3.5\sigma$ or $\pm 4\sigma$ to study the systematic uncertainties
from the requirements on $RM_{\bar{\Sigma}^- p}$ and $RM_{\bar{\Sigma}^- \pi^+}$. The largest
deviations from the nominal values, 0.14\% and 0.16\%, are taken as the systematic uncertainties for
$\mathcal{B}_{\Sigma^+ \to p \pi^0}$ and $\mathcal{B}_{\Sigma^+ \to n \pi^+}$, respectively.

The uncertainties from the MC generators are investigated by varying the decay parameters by $\pm 1\sigma$ of the measured values~\cite{BESIII:2025jxt, BESIII:2023sgt}.
The corresponding efficiency changes are taken as the systematic uncertainties,
which are 0.05\% and 0.04\% for $\mathcal{B}_{\Sigma^+ \to p \pi^0}$ and $\mathcal{B}_{\Sigma^+ \to n \pi^+}$, respectively.
The uncertainties due to the MC statistics are estimated to be
0.05\% and 0.04\% for $\mathcal{B}_{\Sigma^+ \to p \pi^0}$ and $\mathcal{B}_{\Sigma^+ \to n \pi^+}$, respectively.

Adding all systematic uncertainties in quadrature, we obtain the
total systematic uncertainties for the measurement of
$\mathcal{B}_{\Sigma^+ \to p \pi^0}$ and $\mathcal{B}_{\Sigma^+ \to n \pi^+}$ to be
0.44\% and 0.59\%, respectively.

\begin{table*}[htbp]
    \caption{The measured BFs $\mathcal{B}_{\Sigma^+ \to p \pi^0}$, $\mathcal{B}_{\Sigma^+ \to n \pi^+}$, and their ratio, along with
    the PDG values. The ``Difference" marks the differences between the results and the PDG values.
    The first and second uncertainties presented in this work are statistical and systematic, respectively.}
    \label{tab:Results comparison}
    \setlength{\extrarowheight}{1.0ex}
     \renewcommand{\arraystretch}{1.0}
    \begin{center}
        \begin{tabular} {l c c c c }
            \hline \hline
            &~~~~~~ $\mathcal{B}_{\Sigma^+ \to p \pi^0}(\%)$ ~~~~~~& ~~~~~~$\mathcal{B}_{\Sigma^+ \to n \pi^+}(\%)$~~~~~~ & ~~~~~~$\mathcal{B}_{\Sigma^+ \to p \pi^0}/B_{\Sigma^+ \to n \pi^+}$~~~~~~\\
            \hline
            This work       & $49.79 \pm 0.06 \pm 0.22$ & $49.87 \pm 0.05 \pm 0.29$ & $0.9984 \pm 0.0016 \pm 0.0073$    \\
            PDG             & $51.47 \pm 0.30 $         & $48.43 \pm 0.30$          & $1.0628 \pm 0.0090$               \\
            Difference      & $4.4 \sigma$              & $3.4 \sigma $             & $5.5 \sigma$                      \\
            \hline \hline
        \end{tabular}
    \end{center}
\end{table*}

\begin{figure}[htbp]
    \begin{center}
        \begin{overpic}[width = 1.0\linewidth]{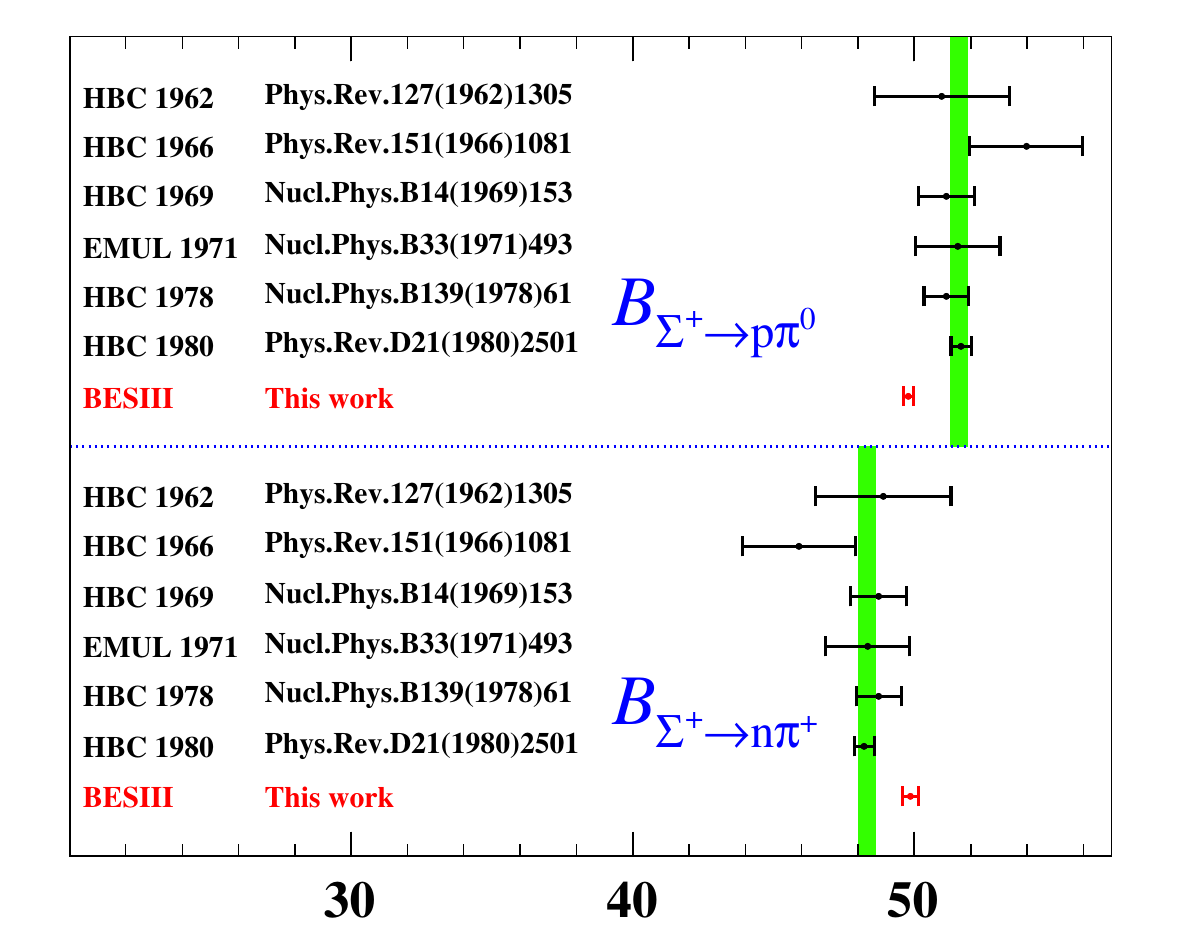}
        \end{overpic}
    \end{center}
    \caption{
        Comparisons of the BFs ($\%$) of $\Sigma^+ \to p \pi^0$ and $\Sigma^+ \to n \pi^+$ with previous experimental
        measurements~\cite{Humphrey:1962zz,Chang:1966zz,Barloutaud:1969tpd,Tovee:1971ga, Nowak:1978au,Marraffino:1980dj}.
        The green bands correspond to the $1\sigma$ limit of the PDG values.
    }
    \label{fig:comparison}
\end{figure}

In summary, using $(10087 \pm 44) \times 10^6$ $J/\psi$ events collected with the BESIII detector, the absolute BFs of $\Sigma^+ \to p \pi^0$ and $\Sigma^+ \to n \pi^+$ are measured for the first time.
The comparison of our results with all the previous measurements is shown in Fig.~\ref{fig:comparison},
where the obtained results are the most precise in the world.
The ratio between $\mathcal{B}_{\Sigma^+ \to p \pi^0}$ and $\mathcal{B}_{\Sigma^+ \to n \pi^+}$ is determined to be
$0.9984 \pm 0.0017 \pm 0.0069$, where the systematic uncertainties associated with the ST yields are canceled out in the calculation.
These results are summarized in Table~\ref{tab:Results comparison}, alongside the PDG values for comparison.
Notably, our measurement deviates from the PDG values~\cite{PDG} by $4.4\sigma$ for $\mathcal{B}_{\Sigma^+ \to p \pi^0}$
and $3.4\sigma$ for $\mathcal{B}_{\Sigma^+ \to n \pi^+}$.
The measured value is somewhat lower than the prediction reported in Ref.~\cite{Wu:2025hnh}. 
Given the model-dependent nature of the calculation, further theoretical improvements will be necessary for a more precise comparison.
As one of the ground-state hyperons, these precisely measured BFs $\mathcal{B}_{\Sigma^+ \to p \pi^0}$ and $\mathcal{B}_{\Sigma^+ \to n \pi^+}$ 
will serve as fundamental inputs for studies of numerous heavier baryon decays.
In addition, the ratio between these two BFs obtained in this Letter exhibits a deviation of $5.5\sigma$ from the PDG value.
Moreover, $\mathcal{B}_{\Sigma^+ \to p \gamma} = (1.19 \pm 0.06) \times 10^{-3}$ has been recalculated
by multiplying our new result of $\mathcal{B}_{\Sigma^+ \to p \pi^0}$ with the PDG average of
$\mathcal{B}_{\Sigma^+ \to p \gamma} / \mathcal{B}_{\Sigma^+ \to p \pi^0}$.
This result differs from the direct measurement by BESIII by $2.9 \sigma$,
which is significantly smaller than the previous reported discrepancy of $4.2 \sigma$~\cite{BESIII:2023fhs}.

Furthermore, the $A$ and $B$ amplitudes of the nonleptonic decays
of $\Sigma^+$ have been updated based on the BFs obtained in this
work, combined with the latest BESIII result of the decay
parameter~\cite{BESIII:2025jxt}. The detailed numerical values can
be found in appendix~\ref{sec:A1}. These more precise measurements will provide a crucial foundation for future studies of CP violation and rare decays in the $\Sigma^+$ sector. Together with the $A$ and $B$
amplitudes for $\Sigma^-$ decay (extracted with the PDG values), the $\Delta I = 1/2$ rule could be
quantitatively tested (in units of $G_F m^2_{\pi^+}$):
\begin{align}
    \label{eq:testA}
    A^{\Sigma^-}_{n\pi^-} - A^{\Sigma^+}_{n\pi^+} + \sqrt{2}A^{\Sigma^+}_{p\pi^0} & = -0.127 \pm 0.014 (\rm this~work) \notag \\
                            & = -0.180\pm 0.063 (\rm PDG), \notag \\
    B^{\Sigma^-}_{n\pi^-} - B^{\Sigma^+}_{n\pi^+} + \sqrt{2}B^{\Sigma^+}_{p\pi^0} & = -2.78 \pm 0.16 (\rm this~work) \notag \\
                            & =  -2.50 \pm 0.76(\rm PDG).
\end{align}
Notably, both quantities deviate from zero by more than $5\sigma$, providing the first compelling evidence for a non-negligible $\Delta I = 3/2$ transition amplitude in $\Sigma$
hyperon decays. The observed violation of the $\Delta I = 1/2$ rule in hyperon decays could help to elucidate the underlying dynamics of isospin symmetry breaking in weak decays. In appendix~\ref{sec:A2}, we present the measured amplitudes of the $S$-wave and $P$-wave, together with the corresponding theoretical predictions from various models. 
Although some models can approximately reproduce parts of the data, none of the existing theoretical frameworks provide a complete description of the observed pattern of amplitudes.
The improved experimental precision reported in this work, therefore, provides important new input for understanding the nonperturbative dynamics
of nonleptonic weak decays in baryon systems.

\textbf{Acknowledgement}

The BESIII Collaboration thanks the staff of BEPCII (https://cstr.cn/31109.02.BEPC) and the IHEP computing center for their strong support. The authors would like to extend thanks to Prof. Jusak Tandean for useful discussions and helpful advice.
This work is supported in part by National Key R\&D Program of China under Contracts Nos. 2023YFA1606000, 2023YFA1606704; National Natural Science Foundation of China (NSFC) under Contracts Nos. 11635010, 11935015, 11935016, 11935018, 12025502, 12035009, 12035013, 12061131003, 12165022, 12192260, 12192261, 12192262, 12192263, 12192264, 12192265, 12221005, 12225509, 12235017, 12342502, 12361141819; 
the China Postdoctoral Science Foundation under Grant No. 2024M753040; 
the Postdoctoral Fellowship Program of China Postdoctoral Science Foundation under Grant No. GZC20241608;
the Chinese Academy of Sciences (CAS) Large-Scale Scientific Facility Program; the Strategic Priority Research Program of Chinese Academy of Sciences under Contract No. XDA0480600; CAS under Contract No. YSBR-101; 100 Talents Program of CAS; The Institute of Nuclear and Particle Physics (INPAC) and Shanghai Key Laboratory for Particle Physics and Cosmology;
Yunnan Fundamental Research Project under Contract No. 202301AT070162;
ERC under Contract No. 758462; German Research Foundation DFG under Contract No. FOR5327; Istituto Nazionale di Fisica Nucleare, Italy; Knut and Alice Wallenberg Foundation under Contracts Nos. 2021.0174, 2021.0299; Ministry of Development of Turkey under Contract No. DPT2006K-120470; National Research Foundation of Korea under Contract No. NRF-2022R1A2C1092335; National Science and Technology fund of Mongolia; Polish National Science Centre under Contract No. 2024/53/B/ST2/00975; STFC (United Kingdom); Swedish Research Council under Contract No. 2019.04595; U. S. Department of Energy under Contract No. DE-FG02-05ER41374.

\appendix

\section{Calculation of $A$ and $B$ amplitudes}
\label{sec:A1}
The amplitude for the spin-1/2 hyperon decay to a spin-1/2 baryon and a pion $B_i \to B_f \pi$ has the form 
\begin{equation}
i \mathcal{M}_{B_i \to B_f \pi} = \bar{u}_{B_f}(A-\gamma_5 B)u_{B_i},
\end{equation}
where $A$ and $B$ are generally complex constants belonging to, respectively, the $S$-wave and $P$-wave components of the transition 
and $\bar{u}_{B_f}$ and $u_{B_i}$ denote the Dirac spinors of the baryons. 
The corresponding decay rate $\Gamma_{B_i\to B_f \pi}$ and the decay asymmetry parameters $\alpha$, $\beta$, 
and $\gamma$ are given by~\cite{PDG}
\begin{equation}
    \Gamma_{B_i \to B_f \pi} = \frac{E_f+m_f}{4\pi m_i}|\vec{\bold p}_f| (|S|^2+|P|^2),
\end{equation}
\begin{equation}
     E_f = \sqrt{m^2_f + \vec{\bold p}^2_f}, ~~S=A, ~~ P  = \frac{|\vec{\bold p}_f|B}{E_f + m_f}, 
\end{equation}
\begin{equation}
       \alpha = \frac{2Re(S^*P)}{|S|^2 + |P|^2}, ~~ \beta = \frac{2Im(S^*P)}{|S|^2 + |P|^2},
    ~~ \gamma = \frac{|S|^2 - |P|^2}{|S|^2 + |P|^2},
\end{equation}
where $\vec{\bold p}_f$ is the three-momentum of $B_f$ in the rest frame of the $B_i$, $m_i$ and $m_f$ 
are the masses of $B_i$ and $B_f$, respectively. 
Then $A$ and $B$ (or $S$ and $P$) amplitudes can be extracted from both the branching fractions and $\alpha$ parameter, 
plus the sign of the $\gamma$ parameter. 

For $\Sigma$ hyperon decays, the nonleptonic amplitudes can be decomposed into isospin components in a notation where superscripts refer to 
$\Delta I = 1/2$, $3/2$~\cite{Donoghue:2022wrw}, 
\begin{align}
    A_{\Sigma^+ \to n \pi^+}  =  \frac{1}{3} A^{(1)}_{\Sigma} -\frac{2}{3} A^{(3)}_{\Sigma} + X_{\Sigma},
\end{align}
\begin{align}
    \sqrt{2} A_{\Sigma^+ \to p \pi^0}  =  -\frac{2}{3} A^{(1)}_{\Sigma} +\frac{4}{3} A^{(3)}_{\Sigma} + X_{\Sigma}, 
\end{align}
\begin{align}
    A_{\Sigma^- \to n \pi^-}  =  A^{(1)}_{\Sigma} + A^{(3)}_{\Sigma}, 
\end{align}
and $X_{\Sigma}$ is of mixed symmetry. 
Therefore, under $\Delta I = 1/2$ limitation: 
\begin{align}
A_{\Sigma^- \to n \pi^-} - A_{\Sigma^+ \to n \pi^+} + \sqrt{2} A_{\Sigma^+ \to p \pi^0}=0,
\end{align}
and similar relations hold for the $B$ amplitudes.

Table~\ref{tab:A_B} shows the current PDG data on the decay parameter $\alpha$, the sign of parameter $\gamma$, the branching fraction $\mathcal{B}$,  
and the extracted $A$, and $B$ amplitudes, which are in the dimensionless units of $G_F m^2_{\pi^+}$. 
In the extraction procedure, final-state interactions are neglected and CP symmetry is assumed, so that $A$, and $B$ are relatively real.

Table~\ref{tab:A_B2} shows the BFs of $\Sigma^+ \to p \pi^0$ and $\Sigma^+ \to n \pi^+$ obtained in this work and 
the decay parameter $\alpha$ for $\Sigma^+ \to p \pi^0$ and $\Sigma^+ \to n \pi^+$ 
from the new BESIII measurements~\cite{BESIII:2025jxt, BESIII:2023sgt} 
(since no CP violation was found, here we take the averaged value for the charge conjugate modes reported in Ref.~\cite{BESIII:2025jxt, BESIII:2023sgt}), and the updated $A$, and $B$ values.

\section{Comparisons of the measured $S$-wave and $P$-wave with various theoretical predictions}
\label{sec:A2}

Table~\ref{tab:Results comparison} shows the measured amplitudes of the $S$-wave and $P$-wave for ${\Sigma^+ \to p \pi^0}$, ${\Sigma^+ \to n \pi^+}$ decays, together with the corresponding theoretical predictions from various models.

\begin{table*}[htbp]
    \caption{The current PDG data on the decay parameter $\alpha$, the sign of parameter $\gamma$, the branching fraction $\mathcal{B}$, 
    and the extracted $A$, and $B$ values.}
    \label{tab:A_B}
    \setlength{\extrarowheight}{1.0ex}
     \renewcommand{\arraystretch}{1.0}
    \begin{center}
        \begin{tabular} {c | c c c | c c}
            \hline \hline
            Decay mode& $\alpha$ & sign of $\gamma$ & $\mathcal{B} (\%)$ & $A$ ($G_F m^2_{\pi^+}$) & $B$ ($G_F m^2_{\pi^+}$) \\
            \hline 
            $\Sigma^+ \to p \pi^0$ & $-0.982\pm0.014 $ & +  & $51.47\pm0.30$   & $ -1.425\pm 0.044$ & $ 11.80 \pm 0.53 $  \\ 
            $\Sigma^+ \to n \pi^+$ & $0.0489\pm0.0026$ & -  & $48.43\pm0.30$   & $ 0.044 \pm 0.002$ & $ 18.56 \pm 0.07 $ \\
            $\Sigma^- \to n \pi^-$ & $-0.068\pm0.008 $ & +  & $99.848\pm0.005$ & $ 1.879 \pm 0.007$ & $ -0.63 \pm 0.07 $  \\
            \hline \hline
        \end{tabular}
    \end{center}
\end{table*}

\begin{table*}[htbp]
    \caption{The updated decay parameter $\alpha$ and the branching fraction $\mathcal{B}$ for $\Sigma^+$, 
    and the extracted $A$, and $B$ values.}
    \label{tab:A_B2}
    \setlength{\extrarowheight}{1.0ex}
     \renewcommand{\arraystretch}{1.0}
    \begin{center}
        \begin{tabular} {c | c c | c c}
            \hline \hline
            Decay mode& $\alpha$ & $\mathcal{B} (\%)$ & $A$ ($G_F m^2_{\pi^+}$) & $B$ ($G_F m^2_{\pi^+}$) \\
            \hline 
            $\Sigma^+ \to p \pi^0$  & $-0.9869\pm0.0019 $ & $49.79\pm0.23$ & $ -1.385\pm 0.008$ & $ 11.80 \pm 0.09 $ \\ 
            $\Sigma^+ \to n \pi^+$  & $0.0506\pm0.0032$  & $49.87\pm0.30$  & $ 0.047 \pm 0.003$ & $ 18.84 \pm 0.07 $ \\
            \hline \hline
        \end{tabular}
    \end{center}
\end{table*}

\begin{table*}[htbp]
    \caption{The measured amplitudes of the $S$-wave and $P$-wave for ${\Sigma^+ \to p \pi^0}$, ${\Sigma^+ \to n \pi^+}$ decays, together with the corresponding theoretical predictions from various models.}
    \label{tab:Results comparison}
    \setlength{\extrarowheight}{1.0ex}
    \renewcommand{\arraystretch}{1.0}
    \begin{center}
        \begin{tabular}{c|cccccccc}
            \hline \hline \text {} & This work  & \textbf{Skyrme\cite{GomezDumm:2000qf}} & \textbf{Skyrme~\cite{Duplancic:2001sm}} &\textbf{CQM~\cite{Tadic:1981fg}} & \textbf{Instanton\cite{Cristoforetti:2004rr}} &\textbf{SU(3)\cite{Zenczykowski:2005cs}} &\textbf{CA-PCAC\cite{Scadron:2000wa}}&\textbf{ChPT\cite{Jenkins:1991bt}} \\
            \hline
            \hline 
            $S(\Sigma^{+} \rightarrow p \pi^0)$ & $-1.385 \pm 0.008$ & -1.09/-0.85 &-1.91 &-1.14 & $-1.56\pm0.28$ &-1.40 &-1.30 &  -1.41 \\
            $P(\Sigma^{+} \rightarrow p \pi^0)$ & $1.177 \pm 0.009$  & 2.18/1.94 &1.15 &0.61 & $0.98\pm0.16$ & 1.26 &1.22 &  0.36 \\
            \hline
            $S(\Sigma^{+} \rightarrow n \pi^{+})$ & $0.047 \pm 0.003$ &  0/0 &0.02 &0 & 0 & 0&0  &-0.09 \\
            $P(\Sigma^{+} \rightarrow n \pi^{+})$ & $1.837 \pm 0.006$ & 1.70/1.35 &1.21 &0.85 & $1.37\pm0.21$ & 1.77 &1.70 &  0.82 \\
            \hline \hline
        \end{tabular}
    \end{center}
\end{table*}

\section{Distributions of $RM_{\bar{\Sigma}^- p}$ and $RM_{\bar{\Sigma}^- \pi^+}$}
\label{sec:A3}
\begin{figure}[htbp]
    \begin{center}
        \includegraphics[width=0.45\textwidth]{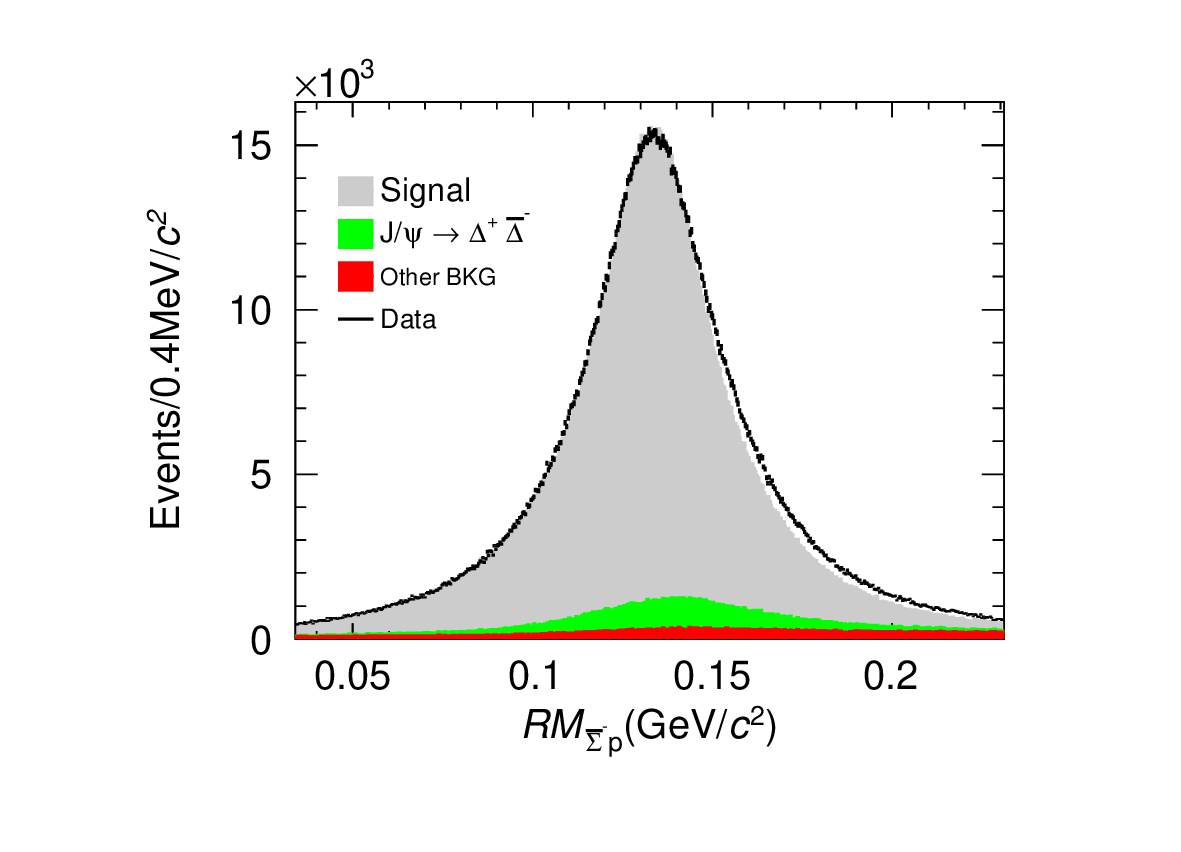}
        \includegraphics[width=0.45\textwidth]{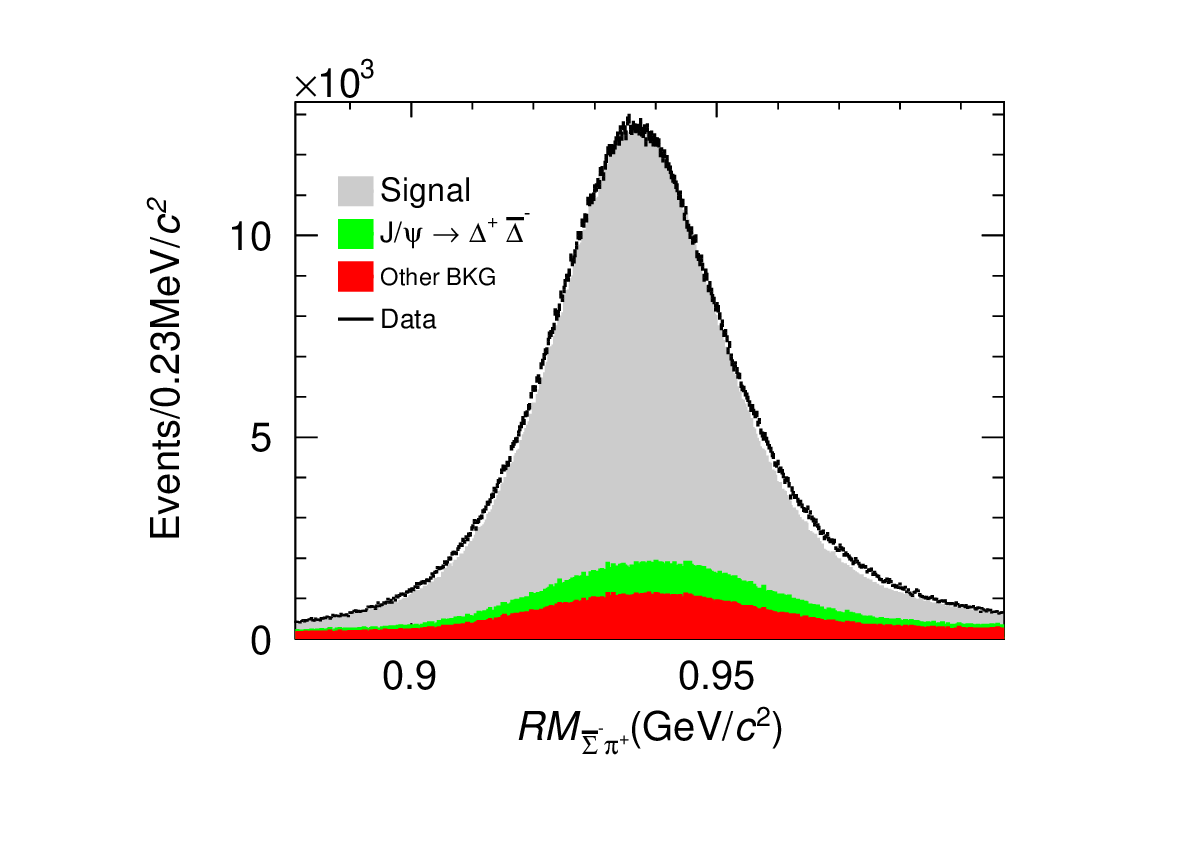}
    \end{center}
    \caption{Distributions of $RM_{\bar{\Sigma}^- p}$ for the $\Sigma^+ \to p\pi^0$ candidate events (left) and $RM_{\bar{\Sigma}^- \pi^+}$ for the $\Sigma^+ \to n\pi^+$ candidate events (right). The black points with error bars are data. The shaded histograms show the expectations from the inclusive MC sample, including the signal (gray), $J/\psi \to \Delta^{+}\bar{\Delta}^{-}$ background (green), and other background processes (red).
    }
    \label{fig:RMSigmaSig}
\end{figure}

Figure~\ref{fig:RMSigmaSig} shows the distributions of $RM_{\bar{\Sigma}^- p}$ for the $\Sigma^+ \to p\pi^0$ candidate events and $RM_{\bar{\Sigma}^- \pi^+}$ for the $\Sigma^+ \to n\pi^+$ candidate events.

%

\end{document}